\documentclass{emulateapj}

\slugcomment{to appear in the Astrophysical Journal}

\usepackage{apjfonts}

\shorttitle{Thermal Stability of White Dwarfs}
\shortauthors{Nomoto et al.}

\begin{document}

\title{Thermal Stability of White Dwarfs Accreting Hydrogen-rich
     Matter and Progenitors of Type Ia Supernovae
}
\author{Ken'ichi Nomoto}
\affil{Department of Astronomy, Graduate School of Science, 
University of Tokyo, 7-3-1 Hongo, Bunkyo-ku, Tokyo 113-0033, Japan}
\email{nomoto@astron.s.u-tokyo.ac.jp}

\author{Hideyuki Saio}
\affil{Astronomical Institute, Graduate School of Science,
    Tohoku University, Sendai, 980-8578, Japan}
\email{saio@astr.tohoku.ac.jp}

\author{Mariko Kato}
\affil{Department of Astronomy, Keio University, Hiyoshi, Kouhoku-ku, 
Yokohama 223-8521, Japan}
\email{mariko@educ.cc.keio.ac.jp}

\and

\author{Izumi Hachisu}
\affil{Department of Earth Science and Astronomy, College of Arts and 
Sciences, University of Tokyo, Komaba, Meguro-ku, Tokyo 153-8902, Japan}
\email{hachisu@chianti.c.u-tokyo.ac.jp}

\begin{abstract}

We revisit the properties of white dwarfs accreting hydrogen-rich
matter by constructing steady-state models, in which hydrogen shell
burning consumes hydrogen at the same rate as the white dwarf accretes
it.  We obtain such steady-state models for various accretion rates
and white dwarf masses.  We confirm that these steady models are
thermally stable only when the accretion rate is higher than $\sim
10^{-7} M_\odot$~yr$^{-1}$.  We show that recent models of ``quiescent
burning'' in the ``surface hydrogen burning'' at a much wider range of
accretion rates results from the too large zone mass in the outer part
of the models; hydrogen burning must occur in a much thinner 
layer.  A comparison of the positions on the HR diagram suggests that
most of the luminous supersoft X-ray sources are white dwarfs
accreting matter at rates high enough that the hydrogen burning shell
is thermally stable.  Implications on the progenitors of Type Ia
supernovae are discussed.

\end{abstract}

\keywords{accretion -- binaries: close -- stars: evolution  --
novae, cataclysmic variables --- supernovae: general -- white dwarfs }

\section{Introduction}

It is widely accepted that progenitors of Type Ia supernovae (SNe Ia)
are mass accreting C-O white dwarfs, although the nature of the
progenitor binary systems is still under debate \citep[e.g.,][]{nom97,
hn00, nom03}.  In the most popular model, the Chandrasekhar mass
model, a C-O white dwarf accretes mass until it grows in mass to $M =
1.38~M_\sun$, then explodes as an SN Ia \citep[e.g.,][]{nm82, nom84,
liv00}.  To find a path to SNe Ia is to find a way for a C-O white
dwarf to increase its central density and temperature to carbon
ignition as the Chandrasekhar limit is approached.  If the mass donor
is a normal star, the stability of the hydrogen burning shell in the
accreting white dwarf is crucial for its evolution.

A hydrogen-shell burning is ignited in an accreting white dwarf when a
certain amount of hydrogen-rich matter is accumulated in the envelope.
The shell burning is unstable to flash if the accretion rate is lower
than a critical rate \citep[e.g.,][]{sug81}.  The flash is stronger if
the white dwarf is cooler and the accretion is slower \citep{pz78,
sat79, fuj82a, fuj82b}.  If the shell flash is strong enough to
trigger a nova outburst, most part of the envelope would be lost from the
system.  Moreover, a part of the original white dwarf matter could be
dredged up and lost in the outburst wind.  Therefore, the white dwarf
mass could decrease after one cycle of nova outburst \citep[e.g.,][]{pri86,
pri95, yar05}, although \citet{tb04} did not find such indication 
in the nova systems.

If the accretion rate $\dot M$ is high enough, on the other hand, the
shell burning is stable and the mass of the white dwarf increases with
accretion.  Thus, the critical accretion rate $\dot M_{\rm stable}$,
above which hydrogen shell burning is stable, is an important physical
value for the evolution of accreting white dwarfs.  (For $\dot M$
being appreciably higher than $\dot M_{\rm stable}$, see \S 5).

Studies on the thermal instability of a thin nuclear-burning shell
began when \citet{sch65} found that helium shell flashes were caused
by a thermal instability of a shell around a C-O core of an evolved
star.  Since then, the stability of a nuclear shell burning in an
accreting white dwarf has been extensively studied in connection with
nova outbursts \citep[e.g.,][and references therein]{sug78}.
\citet{si80} obtained the stability boundary by examining thermal
stability of steady-state models for accreting white dwarfs of various
masses.  The stability boundary, which is $\dot M_{\rm stable} \simeq
10^{-7}M_\odot$yr$^{-1}$ at $M = 0.8M_\odot$ and higher for more
massive white dwarfs, is consistent with fully time-dependent
calculations \citep[e.g.,][]{pz78, sat79, iben82, pri95, yar05}.

Recently, \citet{sta04} published new models of accreting white dwarfs
called ``surface hydrogen burning'' models.  They tried to show that if a
very hot white dwarf after a nova outburst started accreting
hydrogen-rich matter, it developed surface hydrogen burning that
stably converts hydrogen into helium and helium into heavier elements;
then the white dwarf mass could grow to the Chandrasekhar limit.  They
obtained such ``surface hydrogen burning'' models for accretion rates
ranging from $1.6\times 10^{-9}$ to $8.0 \times 10^{-7}
M_\sun$~yr$^{-1}$, and identified these models as supersoft X-ray
sources.  The properties of their models, however, clearly contradict
previous results of evolutionary and steady-state models
\citep[e.g.,][]{iben82, pri95, si80}.

We will show in this paper that the contrasting properties of the
``surface hydrogen burning'' models result from too coarse zoning for
the envelope structure.  We first re-calculate steady-state models for
various accretion rates and white dwarf masses.  If accretion
continues long enough, the mean luminosity of the white dwarf should
approach the steady-state luminosity irrespective of the initial
luminosity.  We then examine the thermal stability of these models to
confirm the results of \citet{si80} using the updated input physics.

In obtaining our steady models, we solve the full stellar structure in
order to avoid the approximations made in the one zone models
\citep{fuj82a}.  We present more detailed numerical models, their
stabilities, and the exact stability boundary values in tabular forms
and formula for a wider range of the white dwarf mass and the
accretion rate compared with \citet{si75, si80}.

We have adopted linear stability analysis mainly because we can avoid
long evolution calculations for many models.  In oder to judge the
stability by evolution calculations as in \citet{iben82}, we have to
adopt sufficiently short time steps, which demands long cpu times.
Also, by a linear analysis we can determine the stability boundary
exactly, while it is difficult for the evolutionary calculations,
because near the stability boundary the growth time is so long that it
would be difficult to distinguish the variation due to thermal
instability from a normal evolutionary change.

Then we examine the ``surface hydrogen burning'' models by \citet{sta04} 
and discuss the reason for the discrepancy with the
previous results.  Our numerical methods are described in \S 2 and our
results are given in \S 3.  In \S 4 we discuss the reason for the
discrepancy between Starrfield et al.'s (2004) models and the previous
results, and compare the loci on the HR diagram of our steady-state
models to those of luminous supersoft X-ray sources.  Implications on
the progenitors of Type Ia supernovae are discussed in \S 5.

\section{Steady-State Models} \label{models}

We have constructed steady-state models for white dwarfs accreting
matter of solar composition.  The steady-state model consists of a C-O
core surrounded by a hydrogen-rich envelope of the solar abundances.
At the bottom of the envelope, hydrogen is burned at the same rate as
the star accretes it; i.e.,
\begin{equation} 
L_{\rm n} = XQ\dot M, 
\end{equation}
where $L_{\rm n}$ is the luminosity due to hydrogen burning, $X$ is
the hydrogen mass fraction in the accreted matter,
$Q=6.4\times10^{18}$erg g$^{-1}$ energy generated when 1 gram of
hydrogen is converted to helium, and $\dot M$ is the mass accretion
rate.  The nuclear reaction rates are taken from \citet{cf88} with the
weak and intermediate screening factors by \citet{gra73}.  The helium
layer and helium shell burning are neglected for simplicity, because
the stability of hydrogen burning in a thin shell is mainly determined
by the structure of hydrogen-rich layer above the hydrogen
burning-shell \citep{sug78}.  Also the helium layer is thin and the
energy release is only 10 \% of hydrogen burning so that the effects
on the stability of hydrogen burning shell are negligible (see
\citet{pac83} for the study of the effect of heat flux from the core).
The compressional heating due to accretion is included in the same way
as in \citet{ksn87}; i.e., the heating rate per unit mass,
$\epsilon_{\rm g}$, is given by

\begin{equation}
\epsilon_{\rm g} = {\dot M\over M}T{ds\over d\ln q},
\end{equation} 
where $s$ is the entropy per unit mass, and $q\equiv M_r/M$ with $M_r$
being the mass included in the sphere of radius $r$.  The chemical
composition of the C-O core is adopted as $(X_{\rm C},X_{\rm O},X_{\rm
Z}) = (0.48,0.50,0.02)$ where $X_{\rm Z}$ denotes the mass fraction of
heavy elements scaled by the solar composition.  Opacity is obtained
from OPAL opacity tables \citep{opal}.

Tables 1 - 4 summarize the properties of several steady state models
for given $M$ and $\dot M$: the mass of the H-rich envelope $\Delta
M_{\rm env}$, $r_{\rm H}$, $T_{\rm H}$, $\rho_{\rm H}$, $P_{\rm H}$,
and $\epsilon_{\rm n}$ at the bottom of the H-rich envelope, and
$L_{\rm n}$, $L$, $R$, and $T_{\rm eff}$ at the surface.

\begin{deluxetable*}{ccccccccc}
\tablecaption{Steady Hydrogen Burning Models ($M = 1.35 M_\odot$)
\label{m135table}
}
\tablewidth{0pt}
\tablehead{
\colhead{$\dot{M}(M_\odot {\rm yr}^{-1}$)}&
\colhead{1.6E-9} &
\colhead{1.6E-8} &
\colhead{1.6E-7} &
\colhead{...} &
\colhead{3.5E-7} &
\colhead{6.0E-7} &
\colhead{...} &
\colhead{*1.6E-7\tablenotemark{a}}
}
\startdata
$\Delta M_{\rm env}/M_\odot$ & 3.7E-7& 2.2E-7& 1.4E-7&...& 1.4E-7& 2.2E-7&...& *1.0E-5 \\
log $r_{\rm H}/R_\odot$ & -2.488 & -2.476 & -2.463 &...& -2.453 & -2.441 &...& *-2.456 \\
log $T_{\rm H}$(K) &   7.67   &   7.82  &   7.98 &...&  8.05  &  8.11  &...& *8.41 \\
log $\rho_{\rm H}$(g cm$^{-3}$) &  2.78  &  2.36   &   1.83  &...&  1.56  &  1.17  &...& *3.22 \\
log $P_{\rm H}$(dyn cm$^{-2}$) & 18.59  &  18.31 &  18.04  &...& 17.98  & 17.98  &...& *19.83 \\
log $\epsilon_{\rm n}$(ergs g$^{-1}$s$^{-1}$) & 9.32 & 10.54 & 11.72 &...& 12.08 & 12.23 &...& *9.34 \\
log $L_{\rm n}/L_\odot$ &  2.048 &  3.049 & 4.049 &...&  4.389  & 4.621 &...& *4.049 \\
log $L/L_\odot$ &   2.072  &   3.073   &  4.077   &...&  4.423  & 4.690 &...& *4.087  \\
log $R/R_\odot$ &  -2.472  &  -2.456  &  -2.421   &...&  -2.376 & -1.746  &...& *-2.334  \\
log $T_{\rm eff}$(K) &  5.516  &  5.758  & 5.992  &...&  6.056 &  5.807  &...&  *5.950 \\
log $\tau_{\rm g}$(s) & 6.86   &  5.94   &  5.64  &...&  stable &  stable  &...&  ...  \\
\enddata

\tablenotetext{a}{The asterisk indicates the quantities for the model
whose $\Delta M_{\rm env}$ is artificially set to be 10$^{-5} M_\odot$
to mimic the ``surface hydrogen burning'' model.}

\end{deluxetable*}

\begin{deluxetable*}{cccccccc}
\tablecaption{Steady Hydrogen Burning Models ($M = 1.25 M_\odot$)
\label{m125table}
}
\tablewidth{0pt}
\tablehead{
\colhead{$\dot{M}(M_\odot {\rm yr}^{-1}$)}&
\colhead{1.0E-9} &
\colhead{1.0E-8} &
\colhead{1.0E-7} &
\colhead{2.1E-7} &
\colhead{...} &
\colhead{4.5E-7} &
\colhead{5.35E-7} 
}
\startdata
$\Delta M_{\rm env}/M_\odot$ & 1.6E-6 & 9.3E-7 & 5.7E-7 & 5.4E-7 &...& 6.8E-7 & 1.1E-6 \\
log $r_{\rm H}/R_\odot$    & -2.306  & -2.298  & -2.282  & -2.273  &...& -2.257  & -2.250 \\
log $T_{\rm H}$(K)   & 7.61  &   7.75  &   7.91   &   7.97  &...&   8.03  &  8.06 \\
log $\rho_{\rm H}$(g cm$^{-3}$) & 2.72  &   2.30  &   1.79   &   1.58  &...&   1.20  &  1.04 \\
log $P_{\rm H}$(dyn cm$^{-2}$) &  18.46 &   18.19 &   17.89  &   17.82 &...&   17.76 &  17.78 \\
log $\epsilon_{\rm n}$(ergs g$^{-1}$s$^{-1}$) & 8.51 & 9.74 & 10.93 & 11.30 &...& 11.54 & 11.61 \\
log $L_{\rm n}/L_\odot$ & 1.845  &  2.845  &  3.845  &   4.167  &...&  4.499 &  4.574 \\
log $L/L_\odot$ &   1.868  &  2.869  &  3.872  &   4.196  &...&  4.538 &  4.628 \\
log $R/R_\odot$ & -2.283 &  -2.267 &  -2.229 &   -2.194 &...&  -2.025 & -1.511 \\
log $T_{\rm eff}$(K) & 5.370 &  5.612 &  5.844 &  5.908 &...&  5.909 &  5.674 \\
log $\tau_{\rm g}$(s) & 7.62  &  6.68  &  6.14 &  7.36  &...&  stable &  stable \\
\enddata
\end{deluxetable*}

\begin{deluxetable*}{cccccccc}
\tablecaption{Steady Hydrogen Burning Models ($M = 1.0 M_\odot$)
\label{m100table}
}
\tablewidth{0pt}
\tablehead{
\colhead{$\dot{M}(M_\odot {\rm yr}^{-1}$)}&
\colhead{1.0E-9} &
\colhead{1.0E-8} &
\colhead{1.0E-7} &
\colhead{1.3E-7} &
\colhead{...} &
\colhead{2.5E-7} &
\colhead{3.6E-7} 
}
\startdata
$\Delta M_{\rm env}/M_\odot$ & 8.0E-6 & 4.7E-6 & 3.2E-6 & 3.2E-6 &...& 3.9E-6 & 6.7E-6 \\
log $r_{\rm H}/R_\odot$ & -2.099 & -2.087 & -2.059 & -2.053 &...& -2.034 & -2.018 \\
log $T_{\rm H}$(K)   & 7.57  & 7.71 & 7.86 & 7.88           &...& 7.93  & 7.96 \\
log $\rho_{\rm H}$(g cm$^{-3}$) & 2.51 & 2.08 & 1.54 & 1.46 &...& 1.20  &  0.97 \\
log $P_{\rm H}$(dyn cm$^{-2}$)  & 18.22  &  17.93  & 17.61  & 17.58  &...& 17.50  &  17.47 \\
log $\epsilon_{\rm n}$(ergs g$^{-1}$s$^{-1}$) & 7.82 &  9.04 & 10.20 & 10.32 &...& 10.59 & 10.66 \\
log $L_{\rm n}/L_\odot$ & 1.845 &  2.845 &  3.845 &  3.959  &...& 4.243  & 4.401 \\
log $L/L_\odot$         & 1.868 &  2.869 &  3.871 &  3.986  &...&  4.274 &  4.439 \\
log $R/R_\odot$ & -2.055 &  -2.026  & -1.946  & -1.923      &...& -1.792  & -1.358 \\
log $T_{\rm eff}$(K) & 5.256 &  5.492 &  5.702 &  5.720     &...&  5.726  &  5.550 \\
log $\tau_{\rm g}$(s) & 8.31 &  7.39  &  7.20  &  8.08      &...&  stable &  stable \\
\enddata
\end{deluxetable*}

\begin{deluxetable*}{cccccccc}
\tablecaption{Steady Hydrogen Burning Models ($M = 0.8 M_\odot$)
\label{m080table}
}
\tablewidth{0pt}
\tablehead{
\colhead{$\dot{M}(M_\odot {\rm yr}^{-1}$)}&
\colhead{1.0E-9} &
\colhead{1.0E-8} &
\colhead{7.5E-8} &
\colhead{...} &
\colhead{1.0E-7} &
\colhead{2.2E-7} 
}
\startdata
$\Delta M_{\rm env}/M_\odot$ & 2.0E-5 & 1.2E-5 & 1.0E-5 &...& 1.0E-5 & 2.0E-5 \\
log $r_{\rm H}/R_\odot$  & -1.990  & -1.973  &  -1.934 &...&  -1.925 &  -1.888 \\
log $T_{\rm H}$(K)   &  7.56 &  7.69  &  7.82  &...&   7.84  &  7.89 \\
log $\rho_{\rm H}$(g cm$^{-3}$) &  2.37  &  1.93  &  1.44  &...&  1.34 &  1.01  \\
log $P_{\rm H}$(dyn cm$^{-2}$) & 18.06  &  17.76  &  17.46  &...&  17.42  &  17.31  \\
log $\epsilon_{\rm n}$(ergs g$^{-1}$s$^{-1}$) & 7.45 &  8.65  &  9.64  &...&  9.76  &  10.04  \\
log $L_{\rm n}/L_\odot$ &  1.845 &  2.845  &  3.720  &...&  3.845   &   4.188 \\
log $L/L_\odot$ &  1.867  &  2.869   &  3.746  &...&  3.871  &   4.218  \\
log $R/R_\odot$ & -1.921  & -1.874 & -1.751  &...& -1.708  &  -1.255  \\
log $T_{\rm eff}$(K) & 5.189  &  5.416  &  5.574 &...&  5.583  &  5.444  \\
log $\tau_{\rm g}$(s) & 8.68  &  7.83  &   8.47  &...&  stable  &  stable  \\
\enddata
\end{deluxetable*}





\section{Thermal Stability of Models} \label{stmodels}

For each steady-state model, we have examined the thermal stability
against a linear perturbation.  Assuming that perturbations to stellar
structure occur without disturbing hydrostatic balance and that they
are spherically symmetric, we have examined thermal stability by
slightly modifying the Henyey-type relaxation code that we use to
compute stellar structure.  As well known, in the Henyey relaxation, a
correction $\delta y_i$ ($i=1,2,3,4$) to a variable $y_i$ is obtained
by solving inhomogeneous linear equations
expressed as                                                     
\begin{equation}                                                 
{\boldmath\mathcal{H}}\delta {\boldmath y} = {\boldmath D}       
\label{eq_heny}                                                  
\end{equation}                                                   
where {\boldmath$\mathcal{H}$} is a $4N\times4N$ matrix with $N$ 
being the number of mesh points, $\delta {\boldmath y}$ a vector 
consists of $\delta y_i$ at each mesh point, and ${\boldmath D}$ 
is a vector with $4N$ components representing the deviation of   
the variables $y_i$ from the correct values.                     
In thermal stability analysis for a stellar model, 
we regard $\delta y_i$ as a perturbation to the variable $y_i$
which satisfies the differential equations for stellar structure 
so that the inhomogeneous terms diminish; i.e., ${\boldmath D}=0$.
We express the temporal variation of perturbed quantities by
$\exp(\sigma t)$ so that we replace the time derivative of entropy
$\partial \delta s/\partial t$ with $\sigma \delta s$.  

We write the modified Henyey matrix as $\overline{\mathcal{H}}$. 
Then, we obtain homogeneous equations 
\begin{equation}                                                  
{\boldmath\overline{\mathcal{H}}}\delta {\boldmath y} = 0,        
\end{equation}                                                    
which govern the perturbation and an eigenvalue $\sigma$.

The eigenvalue $\sigma$ is obtained by searching for a value which
makes the 
matrix $\overline{\mathcal{H}}$                                   
singular.  If there is a positive $\sigma$,
the structure is thermally unstable with a growth rate of $\tau_{\rm
g} = 1/\sigma$.  If all the eigenvalues are negative, on the other
hand, we judge that the stellar model is thermally stable.  The
obtained $\tau_{\rm g}$ and the stability for several models are
summarized in Tables 1 - 4.

\begin{figure}
\epsscale{1.15}
\plotone{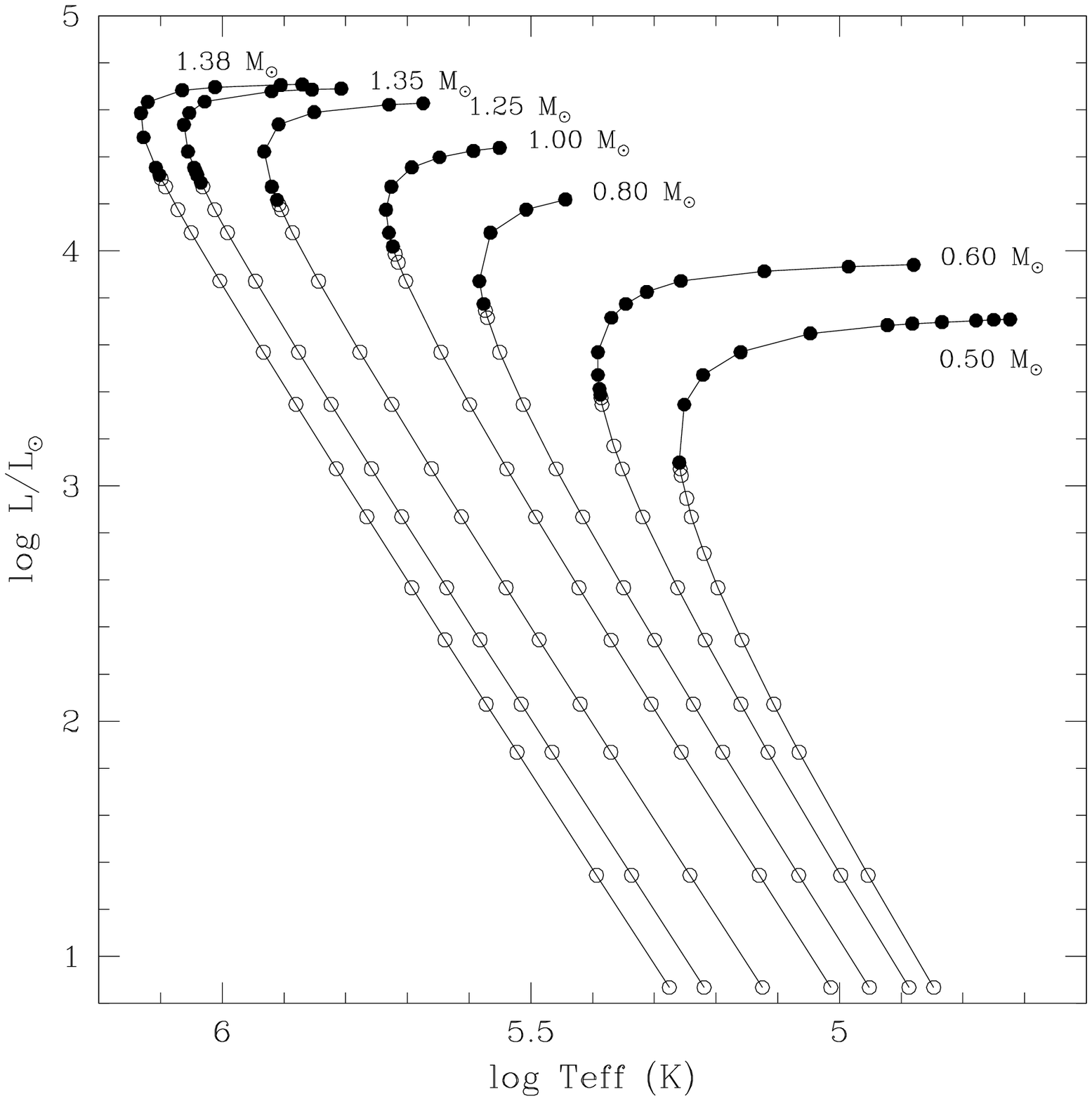}
\caption{
Loci of steady models in the HR diagram.  Each sequence corresponds to
a mass of white dwarf which accretes matter of solar composition at
various rates.  The accretion rate determines the luminosity because
hydrogen is assumed to burn at the same rate of accretion.  Open and
filled circles indicate thermally unstable and stable models,
respectively.
}
\label{hrd}
\end{figure}


Figure~\ref{hrd} shows the steady-state models in the HR diagram
computed for C-O white dwarfs of $M = 0.5 - 1.38 M_{\odot}$ which
accrete hydrogen-rich matter at various rates $\dot M$.  The lowest
luminosity in Figure~\ref{hrd} corresponds to $\dot M = 10^{-10}
M_{\odot}$~yr$^{-1}$.  

As $\dot M$ gets larger along a sequence of models with given mass
$M$, the luminosity asymptotically approaches a limiting value of
$L_{\rm RG}$, which corresponds to the red-giant luminosity determined
by the ``core mass - luminosity'' relation \cite{pac70}.  (Here the
core mass corresponds to the white dwarf mass $M$.)  The exact value
of $L_{\rm RG}$ as a function of $M$ depends on the opacity and
abundance of the accreted matter, so that we give $L_{\rm RG}$ in
Table 5 from our asymptotically obtained values.

As the luminosity approaches $L_{\rm RG}$ the radius of the white
dwarf gets larger.  The open and filled circles indicate thermally
unstable and stable models, respectively.  Models brighter than the
critical luminosity, $L_{\rm stable}$, around the ``knee'' of the
model sequence are thermally stable.

\begin{deluxetable*}{cccccc}
\tablecaption{Stability Boundary for Steady Burning Models
\label{tab5}
}\tablewidth{0pt}
\tablehead{
\colhead{} &
\multicolumn{2}{c}{\underline{~~~~~Stability Boundary~~~~~}} &
\colhead{} &
\multicolumn{2}{c}{\underline{~~~~~Red Giant Boundary~~~~~}}\\
\colhead{$M/M_\odot$}&
\colhead{$\dot{M}_{\rm stable}(M_\odot {\rm yr}^{-1})$}&
\colhead{log $L_{\rm stable}/L_\odot$}&
\colhead{}&
\colhead{$\dot{M}_{\rm RG}(M_\odot {\rm yr}^{-1})$}&
\colhead{log $L_{\rm RG}/L_\odot$}
}
\startdata
   0.5   &    1.65E-8 &  3.08    & ... &    6.9E-8  &  3.71  \\
   0.6   &    3.25E-8 &  3.38    & ... &    1.2E-7  &  3.95  \\
   0.7   &    5.35E-8 &  3.60    & ... &    1.6E-7  &  4.08  \\
   0.8   &    7.75E-8 &  3.76    & ... &    2.3E-7  &  4.23  \\
   0.9   &    1.05E-7 &  3.89    & ... &    2.95E-7 &  4.35  \\
   1.0   &    1.35E-7 &  4.00    & ... &    3.7E-7  &  4.45  \\
   1.1   &    1.65E-7 &  4.09    & ... &    4.4E-7  &  4.53  \\
   1.2   &    1.95E-7 &  4.16    & ... &    5.1E-7  &  4.60  \\
   1.25  &    2.15E-7 &  4.20    & ... &    5.4E-7  &  4.64  \\
   1.30  &    2.35E-7 &  4.25    & ... &    5.8E-7  &  4.67  \\
   1.35  &    2.55E-7 &  4.28    & ... &    6.0E-7  &  4.70  \\
   1.38  &    2.75E-7 &  4.31    & ... &    6.2E-7  &  4.71
\enddata
\end{deluxetable*}


Table 5 summarizes the critical luminosities $L_{\rm stable}$ and
$\dot M_{\rm stable}$, and the limiting luminosities $L_{\rm RG}$ and
corresponding $\dot M_{\rm RG}$.  The critical $\dot M_{\rm stable}$
and $\dot M_{\rm RG}$ are, respectively, approximated by
\begin{equation}
\dot M_{\rm stable} = 3.066 \times 10^{-7} (M/M_\odot - 0.5357)
M_\odot {\rm yr}^{-1},
\end{equation} 
and 
\begin{equation}
\dot M_{\rm RG} = 6.682 \times 10^{-7} (M/M_\odot - 0.4453) M_\odot
{\rm yr}^{-1}.
\end{equation} 
The ratio between these two rates only slightly depends on $M$ as
$\dot M_{\rm RG}/\dot M_{\rm stable}$ = 2.3, 2.5, 2.7, and 3.0 for $M
=$ 1.38, 1.25, 1.0, and 0.7 $M_\odot$, respectively.  The loci on the
HR diagram and the stability of the steady-state models agree with the
results of \citet{si75,si80}.  The above results have recently been
confirmed by \citet{shen07}, who have applied the analytic approach
with the one zone model and examined the metallicity dependence.

\begin{figure}
\epsscale{1.15}
\plotone{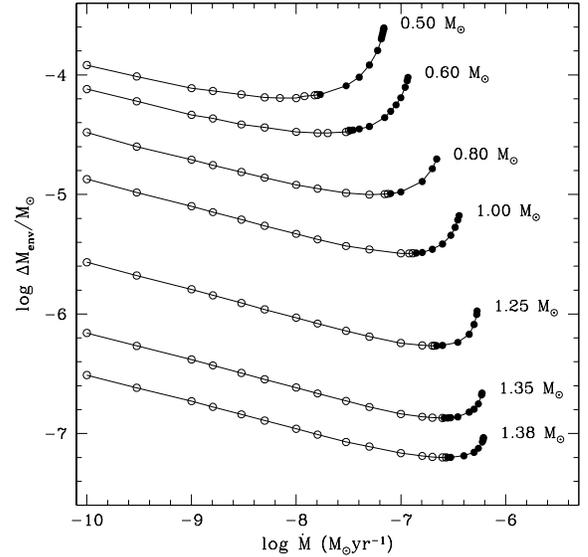}
\caption{
The mass of the hydrogen-rich envelope versus accretion rate for
various white dwarf masses of steady models.  Open and filled circles
indicate thermally unstable and stable models, respectively.
}
\label{envmass}
\end{figure}


Figure~\ref{envmass} shows the hydrogen-rich envelope mass $\Delta
M_{\rm env}$ as a function of $\dot M$ for each mass of the accreting
white dwarf.  The symbols have the same meanings as those in
Figure~\ref{hrd}.  The larger the white dwarf mass, the smaller the
envelope mass for a given accretion rate.  The mass $\Delta M_{\rm
env}$ attains minimum at the accretion rate $\dot M_{\rm stable}$ in
equation (3) corresponding to $L_{\rm stable}$.  For smaller $\Delta
M_{\rm env}$, $T_{\rm H}$ would be too low to give enough $L_{\rm n}$.
For $\dot M > \dot M_{\rm stable}$, the models are thermally stable.

\begin{figure}
\epsscale{1.15}
\plotone{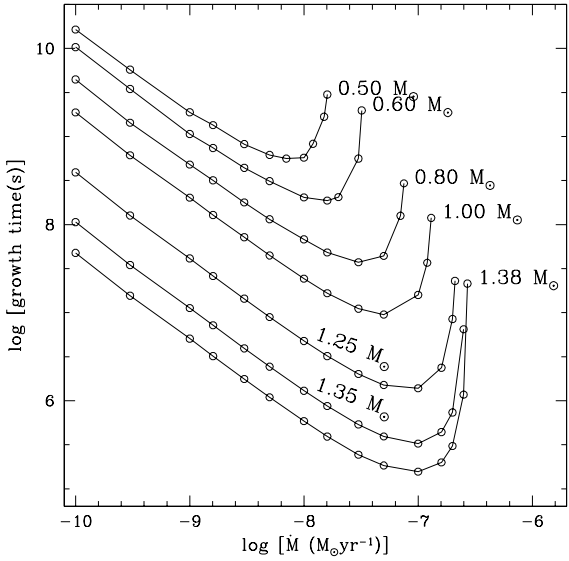}
\caption{
Growth time of the thermal instability of the hydrogen burning shell
is plotted against the accretion rate for each white dwarf mass.
}
\label{growthtime}
\end{figure}


Figure~\ref{growthtime} shows growth time $(=1/\sigma)$ of the thermal
instability of steady-state white dwarf models as a function of the
accretion rate $\dot M$ for each white dwarf mass $M$.  For given $M$,
the growth time decreases as $\dot M$ increases when $\dot M$ is
relatively low.  As $\dot M$ increases, the growth time reaches a
minimum and increases rapidly toward the stability boundary.  For
given $\dot M$, the growth time is longer for smaller $M$.  

To interpret the above results, we follow the description of basic
physics of the stability of nuclear burning in a thin shell as
summarized in \citet{sug78} and \citet{sug81}.  If a temperature
perturbation of $\delta T(M_r) > 0$ is applied to a nuclear burning
shell, it leads to the positive change in specific entropy of $\delta
s(M_r) > 0$ in most cases mainly because of the larger temperature
sensitivity of nuclear reaction rate over radiative loss.  Then
$\delta s(M_r)$ induces the hydrostatic readjustment and the resultant
temperature change is given as:
\begin{equation}
\delta {\rm ln}~T = {1\over c_{\rm g}^*}\delta s,
\end{equation} 
\begin{equation}
{1\over c_{\rm g}^*} = {1\over c_P} + \left({\partial {\rm ln}~T \over
\partial {\rm ln}~P}\right)_s {\delta {\rm ln}~P\over \delta s}.
\end{equation} 
Here $c_P$ and $c_{\rm g}^*$ denote the ordinary thermodynamic specific
heat and the gravothermal specific heat, respectively.  

If $c_{\rm g}^* > 0 (< 0)$, $\delta {\rm ln}~T > 0 (< 0)$ for $\delta
s > 0$ and shell burning is thermally unstable (stable).  Thus the
stability is determined by the hydrostatic readjustment $\delta {\rm
  ln}~P/\delta s (< 0)$ due to expansion (for $\delta s > 0$), which
depends on on the following two factors.

\noindent
(1) {\sl Geometry}: In the extremely thin shell, the pressure at the
bottom of the thin envelope is determined as:
\begin{equation} 
P = {GM^2 (1-q)\over 4 \pi R^4}, 
\label{pressure_q}
\end{equation}
where $R$ is the radius of the white dwarf, and $P$ is determined only
by the column density above the radius $r$.  Therefore, the effect of
expansion for $\delta {\rm ln}~P/\delta s$ is too small to cool the
shell and to stabilize nuclear burning.  This is the main reason for
thin shell burning to be unstable, being the case for low $\dot M$.
For high $\dot M$, entropy at the burning shell is larger, thus
leading to a more extended envelope as seen in Figure~\ref{hrd}.  Then
the effect of hydrostatic readjustment (expansion) is larger and tends
to stabilize nuclear shell burning.

\noindent
(2) {\sl Equation of State}: If electrons are degenerate, $P$ depends
only weakly on $T$, which makes the effect of expansion too small to
stabilize nuclear burning upon $\delta T > 0$.  This is the case for
low $\dot M$ and thus low $s$ at the burning shell.  On the contrary,
for high $\dot M$ and $L$, especially, near their RG values, radiation
pressure is important and its large $T$ sensitivity stabilizes shell
burning (see also \citep{shen07}).

At $L_{\rm stable} < L < L_{\rm RG}$ (or $\dot M_{\rm stable} < \dot M
< \dot M_{\rm RG}$), therefore, these combined effects of radiation
pressure and the extended envelope structure lead to stable burning.
This is why $\dot M_{\rm stable}$ is smaller than $\dot M_{\rm RG}$ by
only a factor of $\sim$ 2.3 - 2.7.  (See \citet{fuj82b} for further
details on the stability of the thin shell approximation models.)

\begin{figure}
\epsscale{1.15}
\plotone{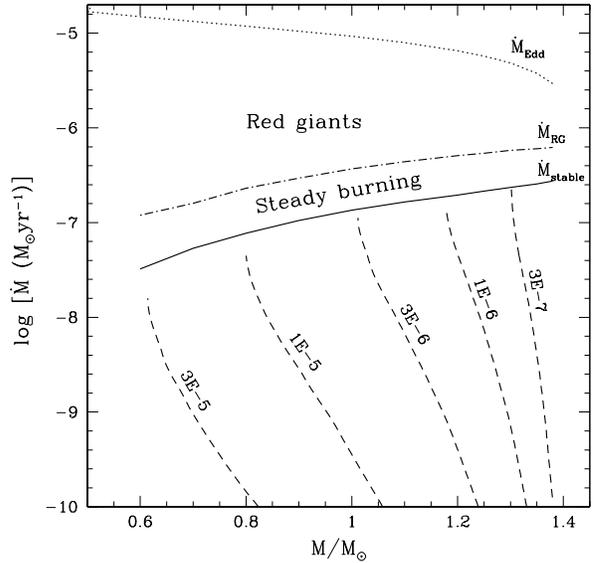}
\caption{
The properties of H-burning shells in accreting white dwarfs are shown
in the white dwarf mass $M$ -- accretion-rate $\dot M$ plane.  If the
accretion rate is lower than the solid line of $\dot M_{\rm stable}$,
H-burning shells are thermally unstable.  A dashed line traces the
locus of the envelope mass $\Delta M_{\rm env}/M_\odot$.  
For given $M$ and $\dot M$, the envelope masses of these steady-state
models are smaller than the envelope masses of the ``ignition'' models
given in Figure~9 in \citet{nm82} because of the higher entropy in the
steady-state models than th `ignition'' models (see text for more
details).
In the area between the solid ($\dot M_{\rm stable}$) and dash-dotted
($\dot M_{\rm RG}$) lines, the H-burning shell burns steadily and the
star is located around the ``knee'' or the horizontal branch on a
locus of steady-state white dwarf models.  Above the dash-dotted line
for $\dot M_{\rm RG}$, the stellar envelope is expanded to a red-giant
size and a strong wind occurs.  Dotted line indicate the Eddington
accretion rate $\dot M_{\rm Edd}$ as a function of $M$.
}
\label{mcmdot}
\end{figure}

Figure~\ref{mcmdot} summarizes the properties of steady-state models
accreting hydrogen-rich matter. The vertical axis is the accretion
rate $\dot M$ and the horizontal axis is the white dwarf mass $M$.
The model properties are classified as follows:

\noindent
(1) The models are thermally unstable in the area below the solid line
to show $\dot M_{\rm stable}$.  The dashed line indicates the loci
where the envelope mass $\Delta M_{\rm env}$ is constant; the value
attached to each dashed line indicates $\Delta M_{\rm env}/M_\odot$.
These envelope masses of the steady state models tend to be smaller
than those obtained by \citet{si75} by a factor of 1.2 - 1.3, but can
be regarded as being consistent in view of the updated opacity in
the present study.

\noindent
(2) In the area above the solid line of $\dot M_{\rm stable}$,
accreting white dwarfs are thermally stable so that hydrogen burns
steadily in the burning shell.

\noindent
(3) Above the dash-dotted line for $\dot M_{\rm RG}$, the accreted
matter is accumulated faster than consumed into He by H-shell burning.
As a result the accreted matter is piled up to form a red-giant size
envelope \cite{nom79}.  The accretion rate is limited by the
Eddington's critical accretion rate $\dot M_{\rm Edd}$ which is
defined as
\begin{equation} 
\dot M_{\rm Edd} = {4 \pi c R\over \kappa}.
\label{eddington}
\end{equation}
For $\kappa = 0.2 (1 + X)$, $\dot M_{\rm Edd}$ is shown by the dotted
line in Figure~\ref{mcmdot}.  Note that, at this critical rate, the
luminosity produced by the gravitational energy release associated
with accretion reaches the Eddington's critical luminosity.

This diagram is similar to Figure~9 in \citet{nm82} except that the
envelope mass in the latter is for the ``ignition'' models.
\citet{nm82} obtained the ``ignition'' model by calculating the
time-dependent evolution of the mass accreting white dwarf with the
Henyey code to find the stage where the nuclear energy generation rate
$\epsilon_{\rm n}$ first exceeds the radiative energy loss rate at the
bottom of the envelope (see \citep{tb04} for the most updated
``ignition'' mass in one zone models).  The envelope mass of
the ``ignition'' model thus determined is larger than the envelope
mass $\Delta M_{\rm env}$ of our steady-state models because of the
lower entropy in the ``ignition'' model for given $M$ and $\dot M$.

The stability of our steady-state models are consistent with the
previous computations for long-term evolutions of accreting white
dwarfs.  For example, \citet{sat79} have found that a $1.2M_\odot$
white dwarf accreting at a rate $1.03\times10^{-7}M_\odot$~yr$^{-1}$
gives rise to repetitive hydrogen shell flashes, while a $1.3
M_{\odot}$ white dwarf accreting at a rate $2.71\times 10^{-7}
M_\odot$~yr$^{-1}$ undergoes stable hydrogen burning.  \citet{pz78}
have also shown that, for a $0.8M_\odot$ white dwarf, the stability
boundary of the hydrogen burning shell is located around $\dot M \sim
10^{-7} M_\sun$~yr$^{-1}$.  Consulting with Figures \ref{envmass} and
\ref{mcmdot}, we can confirm that those evolutionary results agree
very well with our results for steady-state models.

\begin{figure*}
\epsscale{1.0}
\plotone{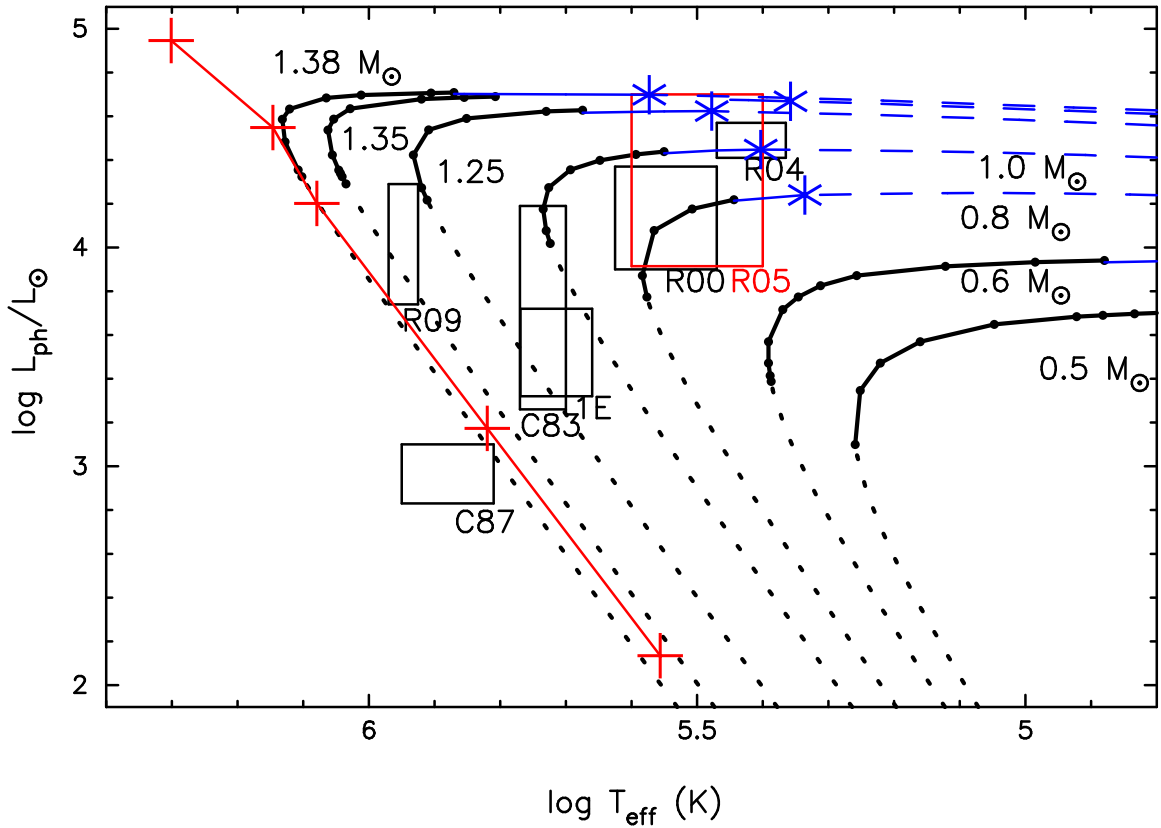}
\caption{
Loci of steady-state models in the HR diagram together with Starrfield
et al.'s (2004) results and positions of several supersoft X-ray
sources.  The filled small circles on the solid line are the same
solutions as in Figure ~\ref{hrd}.  The white dwarf mass is attached
to each curve.  The solid line-parts indicate the stable steady-state
solution, while the dotted line parts correspond to the unstable
solutions as shown in Figure \ref{hrd}.  In the dashed line parts, the
optically thick winds are predicted to blow \citep{kat94}, so that
supersoft X-rays could be absorbed by the wind matter and not be
detected.  The solid line with the crosses show the positions of the
``surface hydrogen burning'' models ($1.35~M_{\odot}$) \citep{sta04}.
Positions of several supersoft X-ray sources are plotted by squares
(taken from Figure 1 of Starrfield et al. 2004), i.e., R09
(RX~J0925.7$-$4758), C87 (CAL87), C83 (CAL83), 1E (1E0035.4$-$7230),
R00 (RX~J0019.8$+$2156), R04 (RX~J0439.8$-$6809).  In addition, the
position of the supersoft X-ray source RX~J0513.9$-$6951 (R05) is also
plotted.
}
\label{hrd2}
\end{figure*}

\section{Discussion}

\subsection{On ``Surface Hydrogen Burning'' Models}

\citet{sta04} wrote that their ``surface hydrogen burning'' models of
mass accreting $1.25 M_\sun$ and $1.35 M_\sun$ white dwarfs are
thermally stable for accretion rates ranging from $1.6\times10^{-9}
M_\sun$~yr$^{-1}$ to $8.0\times10^{-7} M_\sun$~yr$^{-1}$.  The
stability property of their models, however, differs from those of our
steady-state models as well as all other models \citep{si80, fuj82a,
shen07}.  Our models indicate that the hydrogen burning shell in the
$1.35M_\odot$ model is thermally unstable if the accretion rate is
less than $2.5\times10^{-7} M_\sun$~yr$^{-1}$
($2.1\times10^{-7}M_\sun$~yr$^{-1}$ for $1.25M_\odot$)
(Fig.~\ref{envmass}).

\citet{sta04} also wrote that the mass accretion onto the hot white
dwarf just after a nova explosion leads to a stable surface hydrogen
burning.  However, the time-dependent calculations by \citet{pri95}
indicated that the H-accretion onto the hot white dwarfs with the
interior temperatures of $1 - 5 \times 10^7$ K leads to a shell flash
for $M = 0.65 - 1.4 ~M_{\odot}$.

In addition to the difference in the stability property, radii of
Starrfield et al.'s models tend to be smaller than those of our models
(Fig.~\ref{hrd2} below).  In particular, their highest luminosity
model ($M = 1.35M_\odot$ and $\dot M = 8\times10^{-7} M_\odot$
yr$^{-1}$) has a white-dwarf size, while our results indicate that
such a high accretion rate makes the star of a red-giant size
(Fig.~\ref{mcmdot}).

These discrepancies are caused by the extremely coarse zoning adopted
in Starrfield et al.'s computations.  \citet{sta04} adopted a surface
zone mass of $10^{-5}M_\odot$, which is much larger than the entire
envelope mass of the steady-state models of $M = 1.35M_\odot$ and
$1.25M_\odot$ as seen in Figures~\ref{envmass} and \ref{mcmdot}.  This
means that the envelope of the ``surface hydrogen burning'' model is
approximated by a single zone having a single temperature and density.
Furthermore, the ``surface zone'' is much deeper than the realistic
hydrogen-rich envelope of the steady-state model corresponding to the
same white dwarf mass and the accretion rate.

Table 1 compares the two white dwarf models with $M = 1.35M_\odot$
accreting hydrogen-rich matter at a rate of $\dot M = 1.6\times
10^{-7}M_\odot$~yr$^{-1}$.  The steady-state model calculated in the
present study has $\Delta M_{\rm env} = 1.4\times10^{-7} M_\odot$ and
$\log T_{\rm H}$(K) = 7.98 (4th column).  For the model with the
asterisk (7th column), the mass of the hydrogen-rich envelope is
artificially set to be $\Delta M_{\rm env} = 10^{-5} M_\odot$, which is
the same as the ``surface zone mass'' adopted by \citet{sta04}.
According to equation (\ref{pressure_q}) and $T_{\rm H}^4 \propto
P_{\rm H} \propto \Delta M_{\rm env}$, the temperature at the burning
shell ($\log T_{\rm H}$(K) = 8.41) is much higher than those of the
steady-state model.

Such a high temperature is comparable to that of the surface zone of
\citet{sta04}, from which we see the reason why \citet{sta04} obtain
very high temperature at the ``surface zone''.  They treated the
envelope between the region of $\log (1-q) \sim (-5) -(-22)$ by a
single mass zone, while our steady-state models resolve the H-rich
envelope with $\sim50$ mass zones.  Obviously, the zoning adopted by
\citet{sta04} is too coarse to obtain a physically realistic stellar
model.

In the heavy envelope model, the temperature at the hydrogen burning
shell is so high that all accreted hydrogen burns in one typical time
step to compute mass accretion, as \citet{sta04} states ``it takes
less time than the time step ($\sim 2 \times 10^6$ s) for all the
infalling hydrogen to burn to helium in this zone''.  In this case,
the nuclear energy generation rate $\epsilon_{\rm n}$ is determined
not by the temperature-dependent nuclear reaction rate but by the
supplying rate of nuclear fuel as
\begin{equation} 
\epsilon_{\rm n} = {XQ\dot M\over \Delta M_{\rm env}}.
\end{equation}

Despite such a high temperature as $\log T$(K) = 8.41, the energy
generation rate thus determined is $\epsilon_{\rm n} = 2.2 \times 10^9$
ergs g$^{-1}$ s$^{-1}$, which is much lower than the $\beta-$limited
reaction rate of the hot CNO cycle $\epsilon_\beta = 6 \times 10^{13}
(X_{\rm CNO}/0.01)$ ergs g$^{-1}$ s$^{-1}$.  Because $XQ\dot M/\Delta
M_{\rm env}$ is constant, being independent of the temperature, the
nuclear burning is stable; it is also steady as expressed by equation
(1).  In other words, the assumed envelope mass $\Delta M_{\rm env}$
is too large and hence the temperature at the nuclear burning shell is
too high for the mass accretion rates they assumed.  All the accreted
hydrogen-rich matter should have been consumed long before it is
pushed into a layer as deep as $M-M_r\sim10^{-5}M_\odot$
\citep{nar80}.

\subsection{Comparisons with Supersoft X-ray Sources}

Supersoft X-ray sources are suggested to be accreting white dwarfs in
which steady hydrogen burning is taking place \citep[e.g.,][]{vdh92}.
\citet{sta04} argued that the properties of their ``surface hydrogen
burning'' models agree with some of supersoft X-ray sources.  In this
subsection, therefore, we compare our steady burning models and the
``surface hydrogen burning'' models in the HR diagram with several
observed supersoft X-ray sources.


In the HR diagram of Figure~\ref{hrd2}, we show the loci of the steady
burning models for given $M$.  The filled small circles on the solid
lines indicate the thermally stable models in Figure~\ref{hrd}, while
the dotted lines correspond to the unstable solutions. For comparison,
the ``surface hydrogen burning'' models with $M = 1.35M_\odot$
\citep{sta04} are shown by the crosses.  In contrast to the ``surface
hydrogen burning'' models, the sequences of stable models shown by the
solid lines bends rightward because of the increasing radius with
increasing $\dot M$.

The large boxes in Figure~\ref{hrd2} indicate the positions of several
supersoft X-ray sources taken from \citet{sta04}.  The bolometric
luminosities of supersoft X-ray sources are very uncertain because
very soft X-rays are easily absorbed by neutral hydrogen.  For
example, although the luminosity of CAL 87 in LMC is lower than the
$1.38M_\odot$ white dwarf sequence in Figure \ref{hrd2}, the
bolometric luminosity of CAL 87 has been estimated to be a few times
$10^{37}$ergs s$^{-1}$, far above the error box \citep{gre00, gre04}.
Taking into account such large uncertainties in the bolometric
luminosity, we may consider that the loci of luminous supersoft X-ray
sources in the HR diagram are not inconsistent with the loci of
thermally stable models (solid line parts).  We suggest that no
supersoft X-ray source is expected in the dashed line region, because
optically thick winds are predicted to blow in this region
\citep{kat94, kat96} and the matter in the winds would absorb most of
the supersoft X-rays.

We have added one more supersoft X-ray source in the LMC,
RX~J0513.9$-$6951.  \citet*{sch93} estimated the surface temperature
$T_{\rm ph} \sim 30-40$~eV and the total luminosity $L_{\rm ph} \sim
L_{\rm x} \sim 2 \times 10^{38}$~ergs~s$^{-1}$ by a black-body
fitting.  On the other hand, \citet{gan98} obtained $L_{\rm ph} \sim
L_{\rm x} \sim (3-9) \times 10^{37}$~ergs~s$^{-1}$ by their model
atmosphere of the white dwarf.  This is a quasi-regular transient
X-ray source with a relatively short X-ray-on ($\sim 40$~days) and a
long X-ray-off ($\sim 100$~days) states.  \citet{hac03ka} proposed a
transition model, in which the white dwarf expands or contracts
intermittently.  In this model, an optically thick wind occurs only
when the star is expanded, and the X-ray-off state is interpreted as
this expanded phase where the supersoft X-rays are absorbed by the
wind materials.

The existence of such a transient supersoft X-ray source as shown in
Figure~\ref{hrd2} may support our results that there exist envelope
expansions for high mass accretion rates.  Note that in Starrfield et
al.'s models the envelopes do not expand even for higher mass
accretion rates.  This transient X-ray phenomenon is common at least
in the LMC and our Galaxy, because such a system exists not only in
the LMC (in a lower metallicity environment) but also in our Galaxy.
A Galactic supersoft X-ray source, V Sge, which is not shown in
Figure~\ref{hrd2}, is a sister system to RX~J 0513.9-6951
\citep[e.g.,][]{gre98,hac03kb}.

The above comparison shows that the properties of luminous supersoft
X-ray sources are consistent with our steady hydrogen-burning white
dwarf models.  

\section{Concluding Remarks}

We have reinvestigated the properties of accreting white dwarfs by
constructing steady-state models, in which hydrogen shell burning
consumes hydrogen at the same rate as the white dwarf accretes it.  We
have confirmed that these steady models are stable only when the
accretion rate is higher than $\dot M_{\rm stable}$ in equation (3).

Our results contradict the ``surface hydrogen burning'' models by
\citet{sta04} who found quiescent stable hydrogen burning for a wide
range of accretion rates down to $\dot M \sim 10^{-9} M_\odot$
yr$^{-1}$.  We believe we have shown that quiescence of ``surface
hydrogen burning'' results from the too large zone mass ($\sim 10^{-5}
M_\odot$) in the outer part of the numerical models, and that hydrogen
burning must occur in a much more superficial layer ($\sim 10^{-7}
M_\odot$).

We have shown that the positions on the HR diagram of most of the
luminous supersoft X-ray sources are consistent with the white dwarfs
accreting matter at rates high enough for hydrogen shell burning to be
thermally stable.

The narrow range of the accretion rate that leads to the steady and
stable hydrogen burning has been confirmed.  Despite that property,
how the progenitor white dwarfs for SNe Ia could grow their masses to
the Chandrasekhar mass by accreting hydrogen-rich matter has been
discussed in \citet[e.g.,][]{hac96, li97, hac99a, hac99b, lan00,
han04, nom00, nom05}

\acknowledgments

This work has been supported in part by the grant-in-Aid for
Scientific Research (14047206, 14540223, 16540211, 16540219, 
17030005, 17033002, 18104003, 18540231) from the JSPS and MEXT in
Japan.






\begin{thebibliography}{}

\bibitem[Caughlan \& Fowler(1988)]{cf88} 
Caughlan, G.R., \& Fowler, W.A. 1988, Atomic Data and Nuclear Data
Tables 40, 283

\bibitem[Fujimoto(1982a)]{fuj82a} Fujimoto, M.Y. 1982, \apj, 257, 752

\bibitem[Fujimoto(1982b)]{fuj82b} Fujimoto, M.Y. 1982, \apj, 257, 767

\bibitem[Graboske et al.(1973)]{gra73} 
Graboske, H.C., DeWitt, H.E., Grossman, A.S., \& Cooper, M.S. 1973,
\apj, 181, 457

\bibitem[G\"ansicke et al.(1998)]{gan98}
G\"ansicke, B. T., van Teeseling, A., Beuermann, K., \& de Martino,
D. 1998, \aap, 333, 163

\bibitem[Greiner \& van Teeseling(1998)]{gre98}
Greiner, J., \& van Teeseling, A., 1998, \aap, 339, L21

\bibitem[Greiner(2000)]{gre00}
Greiner, J., 2000, New Astr, 5, 137 

\bibitem[Greiner et al.(2004)]{gre04}
Greiner J., Iyudin A., Jimenez-Garate M., Burwitz V., Schwarz R.,
DiStefano R., Schulz N., 2004, in IAU Coll. 194, Compact Binaries in
the Galaxy and Beyond, eds. G. Tovmassian, E. Sion, Rev Mex AA, 20, 18

\bibitem[Hachisu \& Kato (2001)]{hac01}
Hachisu, I., \& Kato, M. 2001, \apj, 558, 323

\bibitem[Hachisu \& Kato(2003a)]{hac03ka}
Hachisu, I., \& Kato, M. 2003a, \apj, 590, 445

\bibitem[Hachisu \& Kato(2003b)]{hac03kb}
Hachisu, I., \& Kato, M. 2003b, \apj, 598, 527

\bibitem[Hachisu \& Kato(2006)]{hac06}
Hachisu, I., \& Kato, M. 2006, \apj, 642, L53

\bibitem[Hachisu et al.(1996)Hachisu, Kato, \& Nomoto]{hac96}
Hachisu, I., Kato, M., \& Nomoto,K. 1996, \apj, 470, L97

\bibitem[Hachisu et al.(1999a)]{hac99a}
Hachisu, I., Kato, M., Nomoto, K., \& Umeda, H. 1999a, \apj, 519,314 

\bibitem[Hachisu et al.(1999b)Hachisu, Kato, \& Nomoto]{hac99b}
   Hachisu, I., Kato, M., \& Nomoto,K. 1999b, \apj, 522, 487 

\bibitem[Han \& Podsiadlowski(2004)]{han04}
   Han, Z., \& Podsiadlowski, Ph. 2004, \mnras, 350, 1301

\bibitem[Hillebrandt \& Niemeyer(2000)]{hn00} 
Hillebrandt, W., \& Niemeyer, J. 2000, \araa, 38, 191 (Errata at
http://arjournals.annualreviews.org/doi/abs/10.1146/annurev.aa.38.010100.200001)

\bibitem[Iben(1982)]{iben82} Iben, I., Jr. 1982, \apj, 259, 244

\bibitem[Iglesias \& Rogers(1996)]{opal} Iglesias, C.A., \& Rogers, F.J.
    1996, \apj, 464, 943

\bibitem[Kato \& Hachisu(1994)]{kat94} Kato, M. \& Hachisu, M. 1994,
   \apj, 437, 802 

\bibitem[Kato(1996)]{kat96} Kato, M. 1996, in Supersoft X-ray Sources, 
    ed. J. Greiner (Heidelberg: Springer), 15

\bibitem[Kawai et al.(1987)]{ksn87} Kawai, S., Saio, H.,
    \& Nomoto, K.  1987, \apj, 315, 229

\bibitem[Langer et al.(2000)]{lan00} 
Langer, N., Deutschmann, A., Wellstein, S., \& H\"oflich, P.  2000,
\aap, 362, 1046

\bibitem[Li \& van den Heuvel(1997)]{li97} Li, X.-D., \& van den
 Heuvel, E.P.J, 1997, \apj, 322, L9

\bibitem[Livio(2000)]{liv00} Livio, M. 2000, in Type Ia Supernovae:
Theory and Cosmology, eds. J. Truran \& J. Niemeyer (Cambridge:
Cambridge Univ. Press), 33

\bibitem[Nariai et al.(1980)Nariai, Nomoto, \& Sugimoto]{nar80}
 Nariai, K., Nomoto, K., \& Sugimoto, D. 1980, \pasj, 32, 473

\bibitem[Nomoto(1982)]{nm82} Nomoto, K. 1982, \apj, 253, 798

\bibitem[Nomoto et al.(1997)Nomoto, Iwamoto, \& Kishimoto]{nom97}
Nomoto, K., Iwamoto, K., \& Kishimoto, N. 1997, Science, 276, 1378

\bibitem[Nomoto et al.(1979)Nomoto, Nariai, \& Sugimoto]{nom79}
Nomoto, K., Nariai, K., \& Sugimoto, D. 1979, \pasj, 31, 287

\bibitem[Nomoto et al.(1984)Nomoto, Thielemann, \& Yokoi]{nom84}
Nomoto, K., Thielemann, F.-K., \& Yokoi, K. 1984, \apj, 286, 644

\bibitem[Nomoto et al.(2000)Nomoto, et al.]{nom00} Nomoto, K., Umeda,
H., Kobayashi, C., Hachisu, I., Kato, M., \& Tsujimoto, T. 2000, in
Type Ia Supernovae: Theory and Cosmology, eds. J. Truran \&
J. Niemeyer (Cambridge: Cambridge Univ. Press), 63 (astro-ph/0003134)

\bibitem[Nomoto et al.(2003)Nomoto, et al.]{nom03} Nomoto, K.,
Uenishi, T., Kobayashi, C., Umeda, H., Ohkubo, T., Hachisu, I., \&
Kato, M. 2003, in From Twilight to Highlight: The Physics of Supernovae, 
eds. W. Hillebrandt \& B. Leibundgut, ESO-Springer Ser. ``ESO
Astrophysics Symposia'' (Berlin: Springer-Verlag), 115
(astro-ph/0308138)

\bibitem[Nomoto et al.(2005)Nomoto, et al.]{nom05} Nomoto, K., Suzuki,
T., Deng, J., Uenishi, T., \& Hachisu, I. 2005, in 1604-2004:
Supernovae as Cosmological Lighthouses, ASP Conf. Ser. 342,
eds. M. Turatto, et al.  (San Francisco: ASP), 105 (astro-ph/0603432)

\bibitem[Paczy\'nski (1970)]{pac70}
  Paczy\'nski, B. 1970, Acta Astr., 20, 47

\bibitem[Paczy\'nski (1983)]{pac83}
  Paczy\'nski, B. 1983, \apj, 264, 282 

\bibitem[Paczy\'nski \& \.Zytkow(1978)]{pz78}
  Paczy\'nski, B., \& \.Zytkow, A.N. 1978, \apj, 222, 604 

\bibitem[Prialnik(1986)]{pri86}
    Prialnik, D. 1986, \apj, 310, 222

\bibitem[Prialnik \& Kovetz(1995)]{pri95}
    Prialnik, D., \& Kovetz, A. 1995, \apj, 445, 789

\bibitem[Schaeidt et al.(1993)Schaeidt, Hasinger, \& Tr\"umper]{sch93}
 Schaeidt, S., Hasinger, G., \& Tr\"umper, J., 1993, \aap, 270, L9

\bibitem[Schwarzschild \& H\"arm(1965)]{sch65} Schwarzschild, M. \&
   H\"arm, R. 1965, \apj, 142, 855

\bibitem[Shen \& Bildsten(2007)]{shen07} Shen, K.J., \& Bildsten,
L. 2007, \apj, in press (astro-ph/0702049)

\bibitem[Sienkiewicz(1975)]{si75} Sienkiewicz, R. 1975, \aap, 45, 411

\bibitem[Sienkiewicz(1980)]{si80} Sienkiewicz, R. 1980, \aap, 85, 295

\bibitem[Sion et al.(1979)]{sat79} 
  Sion, E.M., Acierno, M.J., \& Tomczyk, S. 1979, \apj, 230, 832

\bibitem[Starrfield et al.(2004)]{sta04}
 Starrfield, S., Timmes, F.X., Hix, W.R., Sion, E.M., Sparks, W.M.,
 \& Dwyer, S.J., 2004, \apj, 612, L53

\bibitem[Sugimoto \& Fujimoto(1978)]{sug78}
 Sugimoto, D., \& Fujimoto, M. Y. 1978, \pasj, 30, 467

\bibitem[Sugimoto \& Miyaji(1981)]{sug81}
 Sugimoto, D., \& Miyaji, S. 1981, in IAU Symposium 93, Fundamental
 Problems in Stellar Evolution, ed. D. Sugimoto, D.Q. Lamb, \&
 D.N. Schramm (Dordrecht: Reidel), 191

\bibitem[Townsley \& Bildsten(2004)]{tb04}
 Townsley, D. M., \& Bildsten, L. 2004, \apj, 600, 390

\bibitem[van den Heuvel et al.(1992)]{vdh92}
 van den Heuvel, E. P. J., Bhattacharya, D., Nomoto, K., \& Rappaport,
 S. 1992, \aap, 262, 97

\bibitem[Yaron et al.(2005)]{yar05}
 Yaron, O., Prialnik, D., Shara, M. M., \& Kovetz, A. 2005, \apj, 623, 398

\end{thebibliography}
\end{document}